\pdfoutput=1 
\documentclass[prd,twocolumn,nofootinbib,preprintnumbers,superscriptaddress,showpacs]{revtex4-1}

\usepackage{graphicx}
\usepackage{bm}
\usepackage{amsmath}
\usepackage{lineno}
\usepackage{amssymb}
\usepackage[utf8]{inputenc}
\usepackage[T1]{fontenc}
\usepackage{lipsum}
\usepackage{stmaryrd}
\usepackage{xcolor}
\usepackage{courier}
\usepackage{multirow}
\usepackage{caption,subcaption}
\usepackage{mathtools}

\usepackage{booktabs}

\usepackage[colorlinks,pdfstartview=FitH]{hyperref}
\hypersetup{linkcolor=blue,citecolor=blue,filecolor=black,urlcolor=blue}

\IfFileExists{dsfont.sty}
	{\usepackage{dsfont}
         \let\mathbb=\mathds
         \newcommand{\id}{\mathds{1}}}
	{\typeout{Package dsfont.sty was not found, using alternative macros.}
         \let\mathds=\mathbb
         \newcommand{\id}{\mbox{1 \kern-.59em {\rm l}}}}

\renewcommand\i{\iota}

\newcommand\jet{\mathcal{J}}

\newcommand\lb{\left(}
\newcommand\rb{\right)}

\renewcommand{\part}{{\rm part}}

\newcommand{\be}{\begin{equation}}
\newcommand{\ee}{\end{equation}}
\newcommand{\bes}{\begin{subequations}}
\newcommand{\ees}{\end{subequations}}
\newcommand{\bea}{\begin{eqnarray}}
\newcommand{\eea}{\end{eqnarray}}

\def\nbox#1#2{\vcenter{\hrule \hbox{\vrule height#2in
\kern#1in \vrule} \hrule}}
\def\sq{\,\raise.5pt\hbox{$\nbox{.10}{.10}$}\,}
\def\sqb{\,\raise.5pt\hbox{$\overline{\nbox{.09}{.09}}$}\,}

\begin{document}
\author{Tejes Gaertner}
\email{tejes.gaertner@st-hughs.ox.ac.uk}
\affiliation{Mathematical Institute, University of Oxford, Oxford, OX2 6GG, U.K.}
\affiliation{Department of Physics and Astronomy, University of California, Los Angeles,CA 90095, U.S.A.}
\author{Jared Reiten}
\email{jdreiten@physics.ucla.edu}
\affiliation{Department of Physics and Astronomy, University of California, Los Angeles,CA 90095, U.S.A.}
\affiliation{Mani L. Bhaumik Institute for Theoretical Physics,  University of California,  Los Angeles,  CA 90095, U.S.A.}

\title{Unsupervised learning in the metric space of jets}

\begin{abstract}
In the first part of this work, we demonstrate how the metric space structure induced by the energy mover's distance can be leveraged for the unsupervised tagging of jets according to their progenitor. Namely, we focus on the task of tagging jets initiated by a top quark from a background of jets initiated by light quarks and gluons. By examining the local neighborhood structure of this metric space, we find that the jets of each class populate the landscape in differing densities. This characteristic can be exploited to accurately cluster jets according to their densities through unsupervised clustering algorithms, such as DBSCAN. In the second part of this work, we modify the metric space by reducing the global notion of connectivity down to a local one and, in the process of doing so, modify our distance metric to be that corresponding to geodesics on an underlying graph. We demonstrate how this modification induces regions of both positive and negative values of curvature, which are then exacerbated through a Ricci flow algorithm. Differences in the curvatures averaged over local patches of the new graph metric space then lead to a flow which separates the signal top jets from the background in a fashion that is completely agnostic to any pre-determined jet labels.

\end{abstract}

\maketitle

\section{Introduction}

In recent years, the adoption of machine-learning (ML) techniques by the scientific community, in general, and the particle physics community, in particular, has become increasingly widespread   \cite{Schwartz:2021ftp, Plehn:2022ftl}. On the particle physics front, such tools have found a natural home in the analysis of jets produced by the Large Hadron Collider (LHC) at CERN. Each jet itself is a collimated beam of $\mathcal{O}(10)$ particles, and with each event producing similar orders of such jets, we see such collisions as being one of Nature's many sources of ``Big Data.'' The jet substructure community has fruitfully adapted an enormous number ML and data science techniques to better understand the internal structure of jets \cite{Larkoski:2017jix, Kogler:2018hem, Komiske:2019jim, Marzani:2019hun}. For instance, early on it was realized that jets and their substructure can be naturally visualized and represented in the data structure of an image \cite{Cogan:2014oua} and subsequently passed through deep convolutional neural networks (DCNNs) for classification tasks \cite{deOliveira:2015xxd, Komiske:2016rsd, Chien:2018dfn}. Variational autoencoders (VAEs) have provided a means of compressing the internal features of jets down to a low-dimensional latent representation \cite{Dillon:2020quc, Dillon:2021nxw, Collins:2021pld}. Anomaly detection has seen a great deal of development in efforts to uncover any potential signs of physics beyond the Standard Model (SM) lurking in the vast data sets produced by the LHC  \cite{Fraser:2021lxm, Kasieczka:2021xcg, Buss:2022lxw}. For an extensive list that covers the many applications of ML to particle physics, see \cite{Feickert:2021ajf}. 
Additionally, ML techniques have allowed the community to enhance its understanding of the SM itself---particularly in the identification and analysis of substructure features produced from the decays of heavy particles, such as top quarks. Top-tagging has become an important arena in this domain \cite{Aguilar-Saavedra:2017rzt, Kasieczka:2017nvn, Heimel:2018mkt, Moreno:2019bmu}, for a review, see \cite{Kasieczka:2019dbj}. 

However, the vast majority of the aforementioned applications rely on neural networks that are trained with labeled data, that is machines that ``learn'' through examples where the identity of a particular jet is known, and then using this learned information to accurately classify new instances of unlabeled jets. Accordingly, such studies fall under the realm of supervised learning. While the performance of such neural networks is impressive, as physicists, we would ideally like to understand and interpret what ML techniques are learning---particularly in terms of quantities for which there exists a physical underpinning and from which one can build intuition, as in the spirit of \cite{Faucett:2020vbu, Bogatskiy:2022czk}. In doing so, further applications can dispense of the need for labeled data and learn the physical features impressed upon the data itself. Doing so would allow the community to graduate to the realm of unsupervised learning. An important step in this direction has been in the unsupervised clustering of jets initiated by heavy resonances such as the $W$ and $Z$ bosons, as well as top quarks through use of an attention mechanism on jet images \cite{Mikuni:2020qds}. 

In addition, concepts from optimal transport (OT) have found a natural home in the realm of jet substructure. The so-called Energy Mover's Distance (EMD)  \cite{Komiske:2019fks, Komiske:2020qhg} provides a natural metric on the space of jets, allowing one to associate a ``cost'' for redistributing the substructure of one jet into that of another. A rich geometry results from such a construction, where various familiar event/jet shape observables can be understood as arising from projections onto submanifolds in this abstract space \cite{Komiske:2020qhg}. The richness and utility of this construction has led to the investigation of various extensions, modifications, and alternatives \cite{Cai:2020vzx, Cai:2021hnn}. Significant work has been carried out in order to best visualize this space by embedding it into a lower dimensional space in such a way to preserve the salient features of the true manifold \cite{Park:2022zov}. 

The primary goal of this work is to demonstrate how one can leverage this geometry in order to cluster QCD and top jets in a way that makes no use of labels for jets, i.e. in a completely unsupervised fashion. This is to say, how one can create a top-tagger by accessing only the geometric information afforded by the EMD itself.

This paper is organized as follows. In Sec.~\ref{sec:emd-landscape} we show how QCD and top jets populate their respective metric spaces induced by the EMD. Quantities of interest will be the distribution of distances between jets, the effective dimensionalities of the respective subspaces the jets populate, as well as notions of nearest-neighbor distances. These considerations reveal distinct differences in the geometry of QCD and top jets and lay the groundwork for subsequent sections which work to utilize these differences for the purpose of top-tagging.

In Sec.~\ref{sec:clustering} we demonstrate how the preceding geometric structure can be used to cluster QCD from top jets via the unsupervised density-based clustering algorithm, DBSCAN. This provides us with an effective means of tagging top jets from QCD jets, purely through the use of the underlying geometry of the data. 

In Sec.~\ref{sec:ricci} we demonstrate how QCD and top jets can be separated geometrically through the process of Ricci flow. Ricci flow relies on exacerbating local curvatures in order to separate distinct community structures that live in a weighted graph, and thus, we begin this section by explicitly laying out the mapping of the EMD space to a modified graph metric space. 

Finally, in Sec.~\ref{sec:conclusion} we conclude and provide an outlook for other interesting jet-tagging tasks that could leverage the underlying data geometry to do so in an unsupervised fashion. As top-tagging has by now become a somewhat canonical task and test bed for ML applications, we compare our work to various other ML-based top-taggers in the literature. While our methods achieve competitive accuracies ($\gtrsim 90\%$), they are not state-of-the-art in this particular sense. However, we argue that our methods clearly distinguish themselves as the only ones that are (1) completely unsupervised, (2) rely on $\mathcal{O}(1)$ input parameters rather than the standard, which is often more than $\mathcal{O}(10^5)$, and (3) have a clear physical interpretation.

\section{EMD landscapes of QCD and top jets}
\label{sec:emd-landscape}

The physical intuition underpinning this work is that jets initiated by light quarks and gluons versus those initiated by top quarks have fundamentally different internal substructures---this is a well-known fact in the jet substructure community. The statistical hypothesis from which this work begins is that the labels for the jet progenitors imprint themselves on their respective substructures, and such labels can be inferred or characterized by carefully analyzing ensembles of such substructures. What is to be demonstrated in this work is that such labels reveal themselves to high accuracy through the geometry of the data itself. This section will serve to lay the groundwork for the rest of the paper, as well as convince the reader why such a thing seems plausible. Subsequent sections will provide the quantitative machinery required to convincingly demonstrate this fact.

To begin, we must define the underlying structure of the data from which the aforementioned geometry emerges. Namely, we represent our jets as an unordered set of massless particles, where the $i^{\rm th}$ particle is specified by ($p_{Ti}$, $\eta_i$, $\phi_i$) coordinates. Normalizing each $p_{Ti}$ by the total jet $p_T^{\rm jet}$, $z_i \equiv p_{Ti}/p_T^{\rm jet}$, the jet is then represented as
\begin{align}
\label{eq:pt-cloud}
\mathcal{J} \lb \eta, \phi \rb  = \sum_{i \in \mathrm{jet}} z_i \, \delta \lb \eta - \eta_i \rb \delta \lb \phi - \phi_i \rb \, .
\end{align}
This data type is known as a point-cloud. Given this form, we see (from a statistical viewpoint) that we can treat any given jet as a discrete probability distribution with support on a subset of the 2D Euclidean plane defined by calorimeter cells on which its constituents are deposited after a collision. It is important to note here that a set of pre-processing steps are required to cast a jet in this format \cite{Cogan:2014oua, deOliveira:2015xxd}. These steps consist of rotations, reflections, and translations to get the jet into a standardized form with common origin in $(\eta$-$\phi)$ space defined at the jet's center.

To begin, let us define the indexing set for our data set
\begin{align}
    I \equiv \{0,1, \dots , N\}\,.
\end{align}
In what follows we will consider a data set consisting of QCD and top jets, two thousand of each, all in the point-cloud format of Eq.~(\ref{eq:pt-cloud}), and we will denote this data set by $D$:
\begin{align}
\label{eq:data-set}
    D = \left\{\jet_{f,i} \lb\eta, \phi \rb  \, \big | \, f \in \{\text{QCD, top} \}\,, \, i \in I \right\}\,,
\end{align}
where the first index $f$ runs over the jet flavor labels and $i$ over the indexing set of each ensemble. Our ensembles and labels come from the public top-tagging data set of \cite{Kasieczka:2019dbj}. The jets therein are generated through simulation with Pythia8 \cite{Bierlich:2022pfr} for collisions at $\sqrt{s}= 14$ TeV, such as those at the LHC. Detector effects are modeled by Delphes \cite{deFavereau:2013fsa} paired with use of the ATLAS card. Jets are clustered with the anti-$k_T$ algorithm \cite{Cacciari:2008gp} via FastJet \cite{Cacciari:2011ma}, with a jet radius of $R=0.8$. Each is the leading jet of their respective event. Finally, the transverse momentum range of the data set is $p_T^{\mathrm{jet}} \in [550, 650]$ GeV.

For this section, as well as the remainder of the work, it is important to clarify our use of the jet flavor index. In this present section, we will make ample use of these flavor labels as a means to gain intuition and understanding for the qualitative differences between QCD and top jets as exemplified by use of the EMD. In later sections, this flavor index will only be utilized as the ``truth label'' to assess the accuracy of each unsupervised classification task performed, as the goal of this work is to tag top jets in a completely unsupervised manner, that is, without any reference to labels affiliated with the data.


With our data set $D$ in hand, we compute EMDs between all pairs of jets \cite{Komiske:2019fks, Komiske:2020qhg}, that is we produce a matrix of distances whose $(i$-$j)^{\rm th}$ element will be given the shorthand 
\begin{align}
    \mathrm{EMD}_{ij} \equiv \mathrm{EMD} \lb \jet_i, \jet _j\rb\,.
\end{align}
As $D$ contains jets with transverse momenta $p_T^{\rm jet} \in [550, 650]$ GeV, and this fact is controlled for at the level of the point-cloud representation of Eq.~(\ref{eq:pt-cloud}), this results in the EMD taking on a dimensionless value less than unity. We may interpret this distance then as $\mathrm{EMD} \sim Q / p_T^{\rm jet}$ where $Q$ is the ``traditional'' EMD with units of GeV and $p_T^{\rm jet}$ is in the range specified above. As the dimensionful EMD has the interpretation of the amount of (angular) work required to redistribute the energy of one jet into another, the dimensionless/normalized EMD then can be interpreted as the scale of angular resolution that must be traversed in morphing one jet's point-cloud into another. 

\begin{figure}
\begin{centering}
\includegraphics[width=\linewidth]{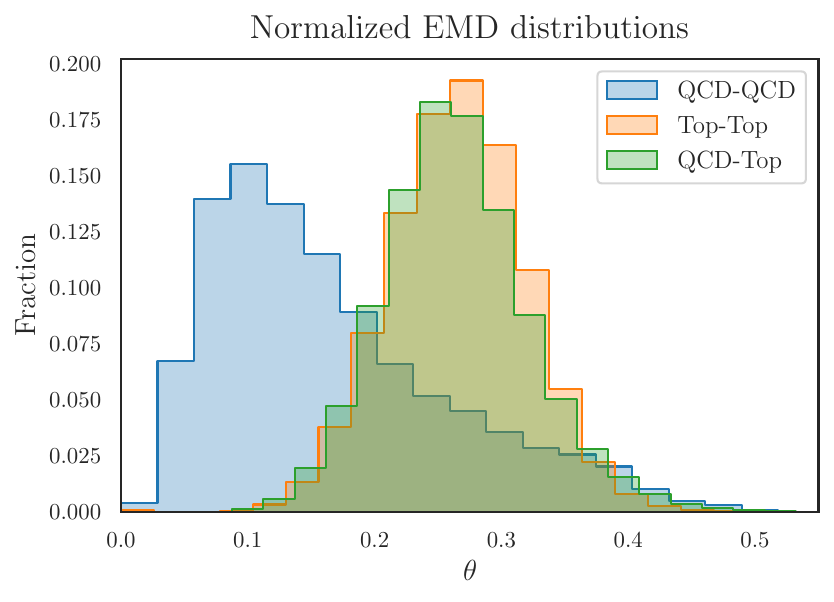}
\end{centering}
 \caption{The three distinct distributions of EMDs affiliated  with our data set of QCD and top jets. The QCD-QCD distribution (blue), the top-top distribution (orange), and the QCD-top distribution (green).}
 \label{fig:emds}
\end{figure}

Our data set $D$ together with the distances between all elements, as computed through the EMD, furnishes us with a metric space $M = (D, \text{EMD})$. The simplest inspection we may perform in the metric space involves visualizing the distribution of EMDs. In doing so, we have three distinct distributions to consider: first, the distances purely between QCD jets (we will refer to this as the ``QCD-QCD'' distribution), second, those purely between top jets (``top-top'') and third, the EMDs between QCD and top jets (``QCD-top''). These distributions are displayed in Fig.~\ref{fig:emds}. This basic figure contains a wealth of information. To begin, looking just at the first moments of the QCD-QCD and top-top distributions, we see that the bulk of QCD jets situate themselves far closer to one another than do top jets---this suggests that the QCD subspace of our EMD metric space is populated at a higher density than is the top subspace. The next striking feature is that the QCD-top distribution nearly overlays the top-top. This is to say that there exists similar spacing between top jets amongst themselves as there does QCD and top jets. Pairing this with the fact that the QCD-QCD distribution is shifted towards lower EMD values relative to the top-top and QCD-top distributions, a significant separation between the QCD jets and the top jets is suggested. Thus Fig.~(\ref{fig:emds}) suggests two qualitative features of the data set: (1) that QCD jets populate a region of higher density than do top jets and (2) these two regions are widely displaced from one another.

Let us further investigate the differing densities in which QCD and top jets populate $M$. To do so, we will define an important class of sets. Sets of this class will be defined for each $\jet_i \in D$,
\begin{align}
\label{eq:emd-set-i}
    \mathrm{EMD}_i \equiv \left\{\mathrm{EMD}_{ij}   \, \big| \, j \in I \,,\, j \neq i \right\}\,,
\end{align}
which is simply the set of EMDs between the jet $\jet_i$ and all other jets in $D$. We use the notation $\mathrm{EMD}_i$ to elicit thoughts of ``taking the $i^{\rm th}$ row'' of the matrix $\mathrm{EMD}_{ij}$, but it is important to note that while $\mathrm{EMD}_i$ contains the elements of the $i^{\rm th}$ row of $\mathrm{EMD}_{ij}$, it is just a set and therefore has no notion of ``location'' for any particular element, as would be captured by the element's column number within the row. We opt for defining this class of sets so that we can order their elements in what follows.

\begin{figure}
\begin{centering}
\includegraphics[width=\linewidth]{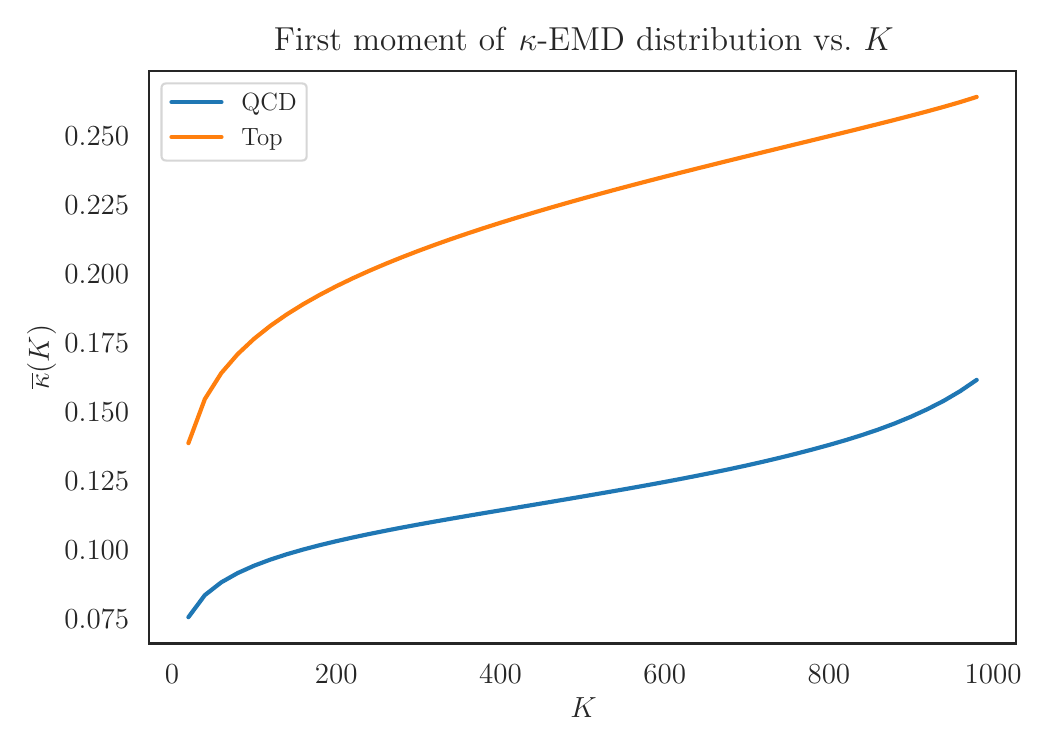}
\end{centering}
 \caption{First moments of the $\kappa$-EMD distributions for QCD (blue) and top (orange) jets, plotted as functions of the nearest-neighbor number $K$.}
 \label{fig:kapppa-emds}
\end{figure}

Next, we will define the $\kappa$-EMD for each jet $\jet_i \in D$ to be the $K^{\rm th}$ order statistic\footnote{For a set of objects in which there exists a natural ordering (as is the case of EMDs as they are non-negative real numbers) the $k^{\rm th}$ order statistic is the $k^{\rm th}$ smallest element, according to the natural ordering, and is denoted $X_{(k)}$. For example, given a set $X = \left\{X_1, \dots, X_n \right\}$, $X_{(1)} = \min X$ and $X_{(n)} = \max X$.} of the set $\mathrm{EMD}_i$:
\begin{align}
\label{eq:kappa-def}
    \kappa _i = \mathrm{EMD}_{i(K)}\,,
\end{align}
which is simply the EMD between $\jet_i$ and its $K^{\rm th}$ nearest-neighbor. We choose this notation to highlight the fact that $\kappa_i$ is a particular matrix element. Thus, given the data set $D$, one can construct distributions in $\kappa$ for each jet type and choice of $K$. Doing so creates distributions in $\kappa$ parameterized by the nearest-neighbor number $K$ and the jet flavor label $f$, which we denote with the notation $p\lb \kappa ; K , f \rb$. An interesting quantity to look at will be the first moments of such distributions as functions of their parameters
\begin{align}
    \overline{\kappa}_f (K)  \equiv \int d\kappa \, \kappa \, p_f\left( \kappa ;K \right)\,.
\end{align}
A plot containing curves for $ \overline{\kappa}_\text{QCD} (K)$ and $ \overline{\kappa}_\text{top} (K)$ is displayed in Fig.~\ref{fig:kapppa-emds}. A striking feature is the clear ordering between the two flavor classes. For every value of $K$, the average distance between a QCD jet and its $K^{\rm th}$ nearest-neighbor is roughly half that of tops. This corroborates the intuition gleaned from Fig.~\ref{fig:emds}, namely that the QCD jets are more tightly packed, and thus occupy a region of higher density in $M$ than do top jets. We will return to this plot when it comes to interpreting the optimal values of input parameters for the DBSCAN algorithm in Sec.~\ref{sec:clustering}.

\begin{figure}
\begin{centering}
\includegraphics[width=\linewidth]{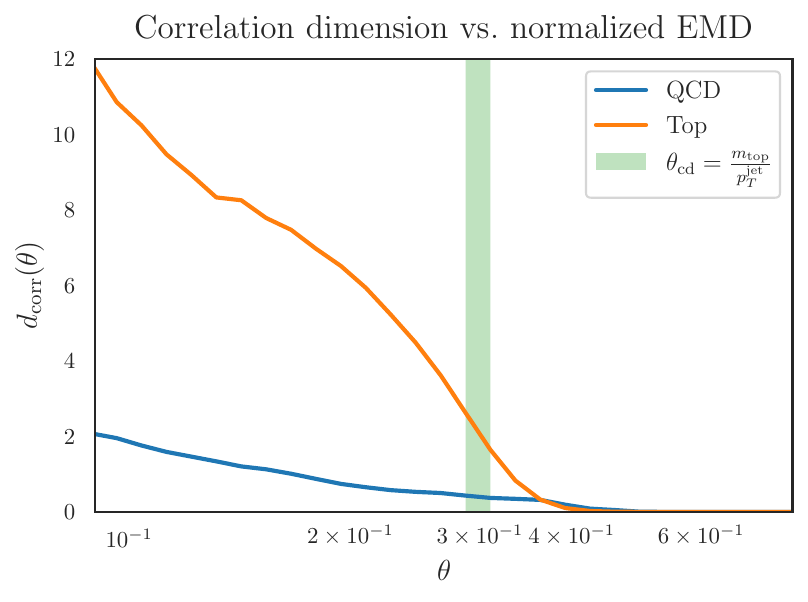}
\end{centering}
\caption{Correlation dimensions for the QCD (blue) and top (orange) subsets of the metric space plotted as a function of the EMD resolution variable $\theta$. The green band corresponds to the range of values taken on by the top's characteristic decay angle $\theta_{\mathrm{cd}}$, whose finite width is due to $p_T^{\mathrm{jet}} \in [550,650 ]$ GeV. This marks the transition region for the resolution of top jets' internal substructures at the level of the geometry of the data.}
 \label{fig:dims}
\end{figure}

The set of features to investigate will be the correlation dimensions \cite{PhysRevLett.50.346, NIPS2002_1177967c, CAMASTRA20032945} for the QCD and top subspaces, as done in the case of $W$-jets in \cite{Komiske:2019fks} as well as both quark and gluon jets in \cite{Komiske:2022vxg}. The correlation dimension is computed as a function of resolution scale $\theta$, according to \cite{Komiske:2019fks}
\begin{align}
\label{eq:corr-dim}
d_{\mathrm{corr}}\left(\theta \right) = \frac{\partial}{\partial \log \theta} \log \sum_{1\leq i < j \leq N}\Theta \left(\mathrm{EMD}_{ij} < \theta \right)\,,
\end{align}
where $\Theta(\cdot)$ denotes the indicator function for the constraint listed as its argument. From a geometric standpoint, smaller $\theta$ resolves local structure while larger $\theta$ resolves more global features of the metric space. Thus from the physics perspective, the local structure of $M$ reflects IR information of the jet, while the global structure captures the UV. We display curves of $d_{\rm corr}$ for the QCD and top subspaces in Fig.~\ref{fig:dims}.

Let us now analyze Fig.~\ref{fig:dims} in detail. We see that the correlation dimension for QCD jets takes the form of a power-law, while that of top jets exhibits a more non-trivial $\theta$-dependence, with the key being an exponential decay in the $\theta \gtrsim 0.3$ region. Both of these observations can be understood as manifestations of the characteristic energy scales for QCD and top jets. 

For QCD jets initiated by light quarks a gluons, the characteristic energy scale is $\Lambda_{\rm QCD}\sim \mathcal{O}(100)$~MeV, which compared to $p_T^{\rm jet} \sim \mathcal{O}(100)$~GeV, is effectively zero. This is to say that the characteristic angular scale for QCD jets is $\theta_{\rm QCD} \sim 0$, or in other words, that QCD jets are approximately scale-free. This lack of definitive angular scale leads to the power-law behavior $d_{\text{corr,QCD}}(\theta) \sim \theta^{-n}$ for some $n > 0$. 

This is in sharp contrast to the top jets, which are distinguished by their characteristic energy scale $m_{\rm top} \sim 173$~GeV. This leads to an angular scale of around $\theta_{\rm cd} \sim m_{\rm top}/p_T^{\rm jet} \in [0.27, 0.31]$. We denote this quantity by $\theta_{\rm cd}$, to allude to the fact that it defines the approximate angular extent of the top's characteristic decay $t \rightarrow q \bar{q}^\prime b$. Looking back to Fig.~\ref{fig:dims} we see that angular scales $\theta > \theta_{\rm cd}$ are too large to resolve any of the substructure features that differentiate top jets from one another, i.e. separate them in the EMD metric space. This gives the top subspace a vanishing correlation dimension. Once we probe the region $\theta \sim \theta_{\rm cd}$, characteristic substructure is resolved, and the complexity of the top jet's population in $M$ becomes manifests itself in a rapid growth in $d_{\text{corr,top}}(\theta)$. In the opposite limit, i.e. $\theta < \theta_{\rm cd}$ we recover an approximate power-law dependence which can be understood as follows. In going far enough below this resolution scale, the top mass is effectively integrated out and we are left with a metric space population pattern characteristic of a scale-free theory like QCD. Physically what is happening is that in probing the metric space at small angular resolution, we are probing neighborhoods of jets who differ only through subtle variations in radiation patterns, generated through QCD evolution, falling collinear to the three (on average) hard prongs marking the characteristic decay pattern of top jets. 

The most significant fact regarding the correlation dimensions of Fig.~\ref{fig:dims} is the striking disparity in their numerical sizes amongst the two jet flavors. At any reasonable angular resolution scale $\theta < R$, where $R$ is the jet radius, QCD jets populate a region of vastly lower dimensionality than do top jets. This provides yet another indication of the high-density population pattern of QCD jets relative to tops in EMD space. For example, consider embedding both subspaces into some larger space that contains both. At every resolution scale, the QCDs are confined to a low-dimensional subspace while the tops are free to diffuse across a larger subvolume.

In total, these three basic considerations regarding the subspaces populated by QCD and top jets indicate that these particular flavor labels imprint themselves in a non-trivial way on the geometry of their resulting metric space. The goal of the following sections is to leverage these geometric features to infer such flavor labels.

\section{Density-based clustering}
\label{sec:clustering}

\subsection{DBSCAN background}
\label{sec:dbscan-business}

In this section, we will give a conceptual overview of the DBSCAN algorithm, as implemented in Scikit Learn \cite{scikit-learn}. Our treatment will closely follow that laid out in the original work \cite{dbscan1}, however instead of describing the general workings, we will tailor things to describe our specific problem at hand. For a more recent treatise, see \cite{dbscan2}.

What makes DBSCAN particularly well-suited to the problem at hand lies in its reliance on two simple parameters that work to quantitatively define a density threshold for elements living in a metric space. Or, to state things in another and equivalent way, what makes our data set so amendable to density-based clustering is the presence of a rather sharp density demarcation in the EMD landscape. As alluded to in the previous sections, when studied individually, sets of QCD- and top-initiated jets populate EMD space with strikingly different densities. While this fact does not guarantee separability in their combined set, it certainly makes such a feat seem promising. Here, we will show that QCD and top jets are indeed separable according to their densities and the requisite density threshold that separates the domains may be extracted from the data in a label-free fashion.

First, let us define several concepts that DBSCAN makes use of. We will make reference to our data set of jets as $D$. The first concept will be that of an $\epsilon$-ball about a given jet, as considered in \cite{Komiske:2020qhg},
\begin{align}
    B_\epsilon \left(\jet_i \right) = \left \{\jet_j \in D  \, \big | \, \mathrm{EMD}_{ij} \leq \epsilon \right\} \,,
\end{align}
that is, the set of jets within an EMD of $\epsilon$ from the given jet $\jet_i$. The next point has to do with two qualitatively different types of points belonging to a cluster, namely what are referred to as \textit{core} and \textit{border} points \cite{dbscan1}. A core point is defined as one whose $\epsilon$-ball has a cardinality bounded below by some value, $\mu$. This is to say that the jet $\jet_i$ is a core point if
\begin{align}
\label{eq:core-point}
    \left| B_\epsilon \left(\jet_i \right) \right| \geq \mu\,.
\end{align}
This bound is fixed in order to contrast core points from border points, where one can easily picture a point lying on the border of a cluster to have fewer jets surrounding it, that is $\left| B_\epsilon \left(\jet_i \right) \right| < \mu$. Next, the jet $\jet_B$ is said to be \textit{directly density-reachable}, with respect to $(\epsilon, \mu)$, by the jet $\jet_A$ if $\jet_A$ is a core point and $\jet_B \in B_\epsilon \lb \jet_A \rb$. If $\jet_B$ then fails to meet the core point condition itself, it is a border point. Thus, we see that the basic notions of core and border points are defined with respect to the parameter pair $(\epsilon, \mu)$, and therefore in what follows, we will consider all further notions regarding the connectivity of clusters in our metric space to be defined with respect to this pair. 

Jets $\jet_A$ and $\jet_B$ are said to be \textit{density-reachable} if there exists a sequence of jets $\lb \jet_1, \jet_2, \dots, \jet_K \rb$ such that $\jet_1 = \jet_A$, $\jet_K = \jet_B$ and $\jet_{i+1}$ is directly density-reachable from $\jet_i$. Jets $\jet_A$ and $\jet_B$ are then said to be \textit{density-connected} if there exists a jet $\jet_C$ from which both $\jet_A$ and $\jet_B$ are density-reachable.

DBSCAN then defines a \textit{cluster}, $C\subset D$ according to the following two conditions: (1) for all $\jet_A$ and $\jet_B$, if $\jet_A \in C$ and $\jet_B$ is density-reachable from $\jet_A$, then $\jet_B \in C$. (2) for all $\jet_A, \jet_B \in C$, $\jet_A$ is density-connected to $\jet_B$. 

Lastly, in contrast to a cluster, DBSCAN defines the \textit{noise}, $\mathrm{noise} \subset D$, to be that which belongs to no cluster, for example, suppose $C_1,\dots, C_K$ are clusters of $D$, then 
\begin{align}
\mathrm{noise} = \left\{\jet \in D  \, \big| \, \jet \notin C_i\,, \,\, \forall i \in \{1,2,\dots, K \}\right\}\,.    
\end{align}
To restate intuitively, the noise set is the set of jets where none of its interior points satisfy the core point condition of Eq.~(\ref{eq:core-point}), which is to say that these points fall below the density threshold established by the parameter pair $(\epsilon, \mu)$.

\subsection{Unsupervised extraction of input parameters}
\label{sec:dbscan-parameters}

\begin{figure}
\begin{centering}
\includegraphics[width=\linewidth]{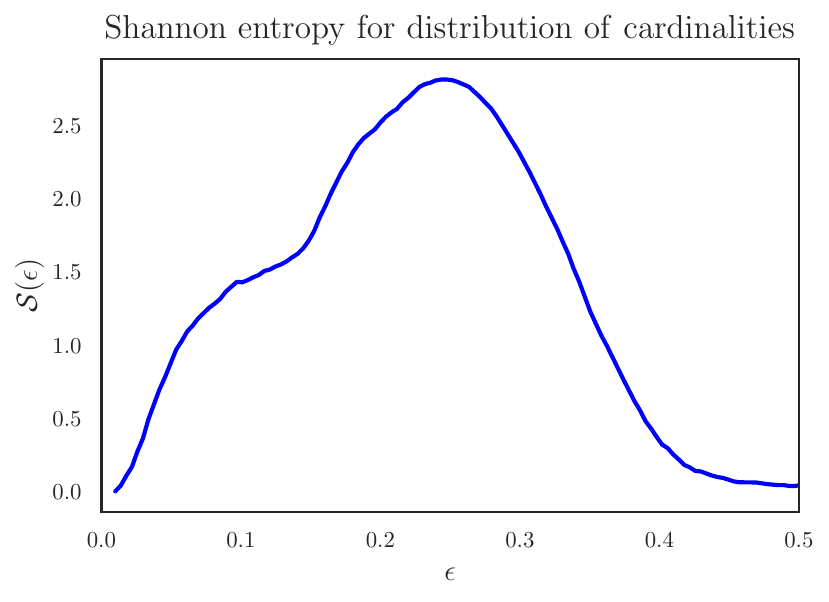}
\end{centering}
 \caption{The Shannon entropy of the cardinality distribution as a function of ball radius $\epsilon$. The location of the local minimum at $\epsilon \sim 0.1$ determines the ideal value of the \texttt{eps} parameter to be used by DBSCAN.}
 \label{fig:entropy}
\end{figure}

\begin{widetext}

\begin{figure}
\centering 
\includegraphics[width=0.45\linewidth]{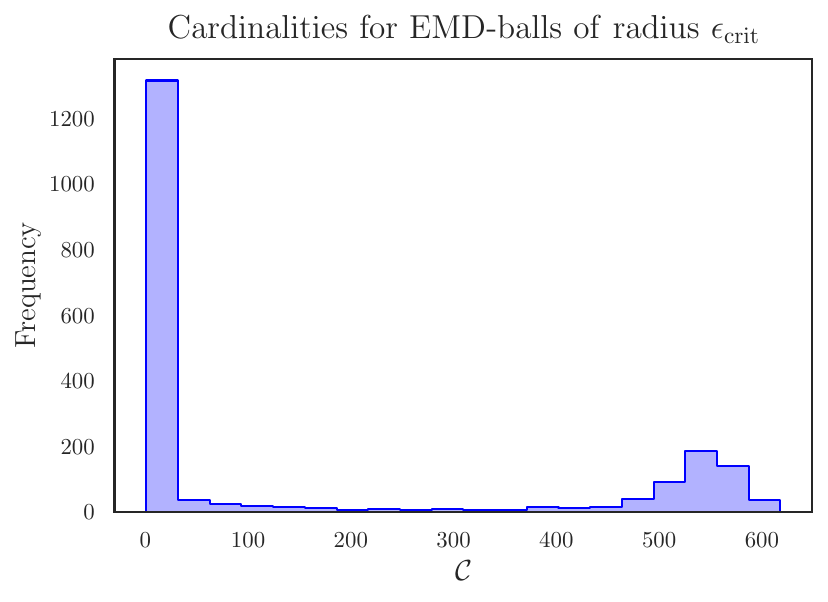}
\hspace{0.5cm}
\includegraphics[width=0.45\linewidth]{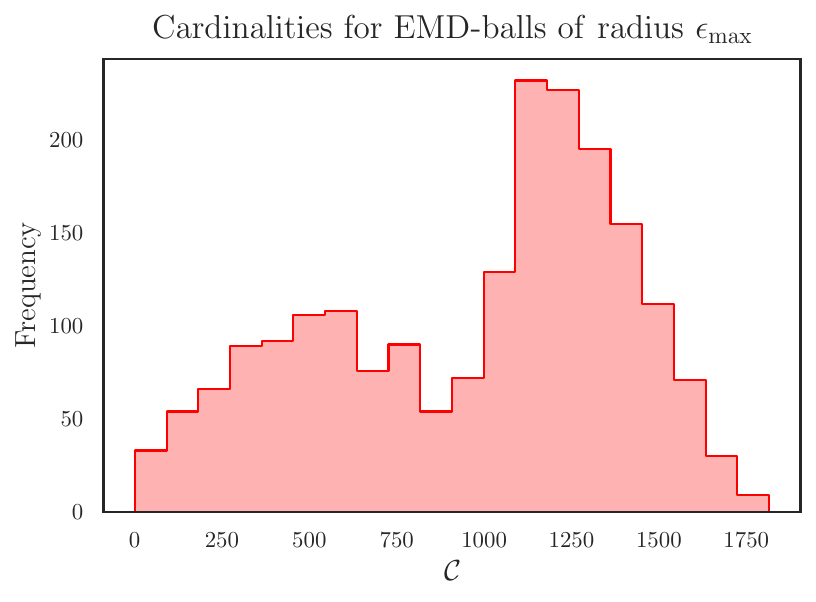} 
\caption{Cardinality distributions corresponding to the two extremum points. (Left) Distribution corresponding to $\epsilon_{\mathrm{crit}} \sim 0.1$, where a well-separated bimodal structure is clear. (Right) Distribution corresponding to $\epsilon_{\mathrm{max}} \sim 0.5$, where the bimodal structure is smeared to a large degree.}
\label{fig:eps-min-samples}
\end{figure}

\end{widetext}

In the previous section, we outlined how DBSCAN defines points based on neighborhoods, and then forms clusters based on the connectivity of such neighborhoods. All definitions were ultimately made with respect to the parameter pair $(\epsilon, \mu)$. This pair forms the basic user-input to the DBSCAN algorithm---using the notation of Scikit-learn \cite{scikit-learn}, $(\epsilon , \mu)\rightarrow (\texttt{eps}, 
 \, \texttt{min\_samples})$. Now, as the goal of this work is to perform top-tagging in an unsupervised way, we need a method to determine $(\epsilon , \mu)\rightarrow (\texttt{eps}, 
 \, \texttt{min\_samples})$ without the use of any labels on our data set. In what follows, we describe a particular procedure for doing so.
 
Our procedure is rather simple. First, we consider the entire unlabeled data set consisting of an equal mixture of QCD and top jets. We construct $\epsilon$-balls about each jet and then compute the cardinality of each ball. The resulting cardinality then ends up being dependent on both $\epsilon$ and the particular jet $\jet$:
\begin{align}
    \mathcal{C}\left( \epsilon, \jet \right) \equiv \left| B_\epsilon \left(\jet \right) \right|\,.
\end{align}


Next, we collect the cardinalities $\forall \jet \in D$ and bin them into a histogram, normalizing by the total number of jets $N+1 = |D|$. We thus consider the resulting histogram to be a discrete distribution in the random variable $\mathcal{C}$, whose stochasticity is derived from that of the underlying set of jets forming $D$, and is parameterized by the variable $\epsilon$. This is to say, that for each value of $\epsilon$, we have a probability distribution $p \left(\mathcal{C} ; \epsilon \right)$, where, despite using ``continuum'' notation, we keep in mind that $\mathcal{C}$ is indeed discrete. To state in other words, the EMD scale $\epsilon$ defines a one-parameter family of cardinality distributions $p \left(\mathcal{C} ; \epsilon \right)$.

Lastly, we consider the one-parameter family of Shannon entropies \cite{10.5555/1146355} resulting from our cardinality distributions
\begin{align}
    \mathcal{S}(\epsilon) = - \sum_{\mathcal{C}} p \left(\mathcal{C}; \epsilon \right) \log p \left(\mathcal{C}; \epsilon \right)\,.
\end{align}
The Shannon entropy quantifies the number of bits (if one uses log base 2) required to specify the distribution. From a more qualitative standpoint, the entropy is a proxy for how much ``uncertainty'' lies in the random variable $\mathcal{C}$. This being the case, we can immediately point understand two special limits. First, we have that 
\begin{align}
    \lim _{\epsilon \rightarrow 0} \mathcal{S}(\epsilon) = 0\,,
\end{align}
since for zero ball-radius, all balls have a cardinality of 1, hence resulting distribution has all its mass contained in a single bin. Similarly, we have that
\begin{align}
    \lim _{\epsilon \rightarrow R} \mathcal{S}(\epsilon) = 0\,,
\end{align}
since $R \gtrsim \max \left\{\mathrm{EMD}_{ij} \, \big | \, i,j \in \{0, 1, \dots, N\} \right\}$ is the maximum angular distance scale occurring in $D$, for in this scenario all balls have a cardinality $n$, so the entirety of the distribution's mass lies in the bin containing $n$. So, since $\mathcal{S} \rightarrow 0$ at both endpoints and $\mathcal{S} \geq 0$, the mean value theorem tells us $\mathcal{S}$ has a global maximum for some value of $\epsilon$. Now, the scale at which $\mathcal{S}$ reaches its global maximum is the scale at which $p \left(\mathcal{C} ; \epsilon \right)$ is the closest it comes to a uniform distribution, which is to say the furthest it comes to any sense of bimodality. 

It turns out that the closest approach to bimodality in $p \left(\mathcal{C} ; \epsilon \right)$ occurs at the $\epsilon$-value for which $\mathcal{S}$ achieves a saddle-point. As $\epsilon$ increases above zero, bin content from from 1 begins to redistributes itself to populate higher values of $\mathcal{C}$---doing so leads to a sharp increase in $\mathcal{S}$. As this continues, bins of higher and higher cardinality are populated, which extends the support set of $p \left(\mathcal{C} ; \epsilon \right)$. Eventually, this rate slows down and reaches a local extremum, as depicted in Fig.~\ref{fig:eps-min-samples}. At this point, our distribution is the approximately bimodal, with a large peak at $\mathcal{C} = 1$ and a localized Gaussian hump near $\mathcal{C} \approx 550$. We refer to this critical value as $\epsilon_{\rm crit}$. As we move into the region $\epsilon \gtrsim \epsilon_{\rm crit}$, more and more bin content from 1 bleeds over into the ``valley'' separating the two peaks, hence lessening their respective levels of localization. This occurs until we reach $\epsilon_{\rm max}$ at which the global max of $\mathcal{S}$ occurs, where the dual-peak structure is most obscured. After this point, the maximally-spread distribution starts localizing/collapsing to the single peak at $n$.

We find that $\epsilon_{\rm crit} \approx 0.1$ and we can understand the emergence of this scale as follows. Consider the EMD distributions for QCD and top jets separately, as displayed in Fig.~\ref{fig:emds}. We find that, in this plot, $Q/p_T \sim \epsilon_{\rm crit}$ marks the lower-tail region of the top distribution, this is to say that there only exist QCD jets with inter-jet distances below $\epsilon_{\rm crit}$. Once we reach this threshold, top jets start coming into the mix. Therefore, $\epsilon_{\rm crit}$ provides a natural ball-radius to assess QCD densities. Displayed in the (b) panel of Fig.~\ref{fig:eps-min-samples} is the distribution $p \left(\mathcal{C} ; \epsilon_{\rm crit} \right)$, where the first bin is expected to be largely populated by top jets, and higher bins, particularly those for  $\mathcal{C} > 300$, are expected to be QCD jets, for these jets must be very dense. It is worth stressing again that while our hypothesis of the labels corresponding to bin contents is inspired by the analysis of previous sections, this entire procedure makes no use of labels. On general grounds, the saddle-point of $\mathcal{S}(\epsilon)$ is a natural scale to emerge, and inspection of $p \left(\mathcal{C} ; \epsilon_{\rm crit} \right)$ reveals a rather striking hint at the presence of two clusters. The vast valley between peaks makes the choice of $\mathcal{C}$ cut quite natural as well. Let us call the midpoint of the valley separating the two peaks $\mathcal{C} = \mu_{\rm crit}$ just to pair with $\epsilon_{\rm crit}$.

Thus, this simple procedure of examining critical EMD value for the one-parameter family of Shannon entropies resulting from the one-parameter family of cardinality distributions provides us with natural input parameters for DBSCAN in a completely unsupervised way. Thus we take
\begin{align}
\label{eq:dbscan-vals}
    (\texttt{eps}, \, \texttt{min\_samples}) = (\epsilon_{\rm crit}, \mu_{\rm crit})\,,
\end{align}
so that DBSCAN defines a core point of a cluster to be
\begin{align}
\label{eq:dbscan-core}
    \left| B_{\epsilon_{\rm crit}} \left(\jet_{\rm core} \right) \right| \geq \mu_{\rm crit}\,.
\end{align}
Interestingly, based on our knowledge from the previous section, we see that by-and-large only QCD jets will meet this criterion, and thus we can expect DBSCAN form a single cluster of QCD jets and classify the top jets as noise. This is natural for our binary data set, as only a single density threshold need exist to partition our data.\footnote{If we had additional jet types, say $W$- or $Z$-jets, occupying the metric space in differing densities, our method could be applied using in successive steps, still using DBSCAN. Alternatively, one could in principle cluster the varying densities using a more sophisticated clustering algorithm, such as HDBSCAN \cite{McInnes2017}. We leave such investigations to future work.}

\subsection{Top-tagging results from density-based clustering}
\label{sec:clustering-results}

Insertion of the EMD matrix for our data set and the application of the parameter values as determined in Eq.~(\ref{eq:dbscan-vals}) to DBSCAN is the full extent of the initialization required for our clustering task. With these specifications, DBSCAN outputs the identification of one set identified as a cluster, which recalling our discussion in Sec.~\ref{sec:dbscan-business}, means that one subset of our data meets the criteria of density-reachability and density-connectivity according to Eq.~(\ref{eq:dbscan-core}), while the other subset does not meet these constraints and is thus identified as noise. Our analysis in Sec.~\ref{sec:emd-landscape} then clearly identifies QCD jets as those forming the cluster with tops making up the noise. The accuracy of this clustering is then assessed through the use of the $F_1$ score, whose value is determined via
\begin{align}
    F_1 = \frac{2 \mathrm{tp}}{2 \mathrm{tp} + \mathrm{fp} + \mathrm{fn}}\,,
\end{align}
where tp denotes a true positive, fp a false positive, and fn a false negative. Such is a standard metric for the assessment of the accuracies of unsupervised clustering tasks. The accuracy we obtain is
\begin{align}
    F_1\bigg|_{\text{EMD+DBSCAN}} = 0.9003\,.
\end{align}

\begin{figure}
\centering 
\includegraphics[width=\linewidth]{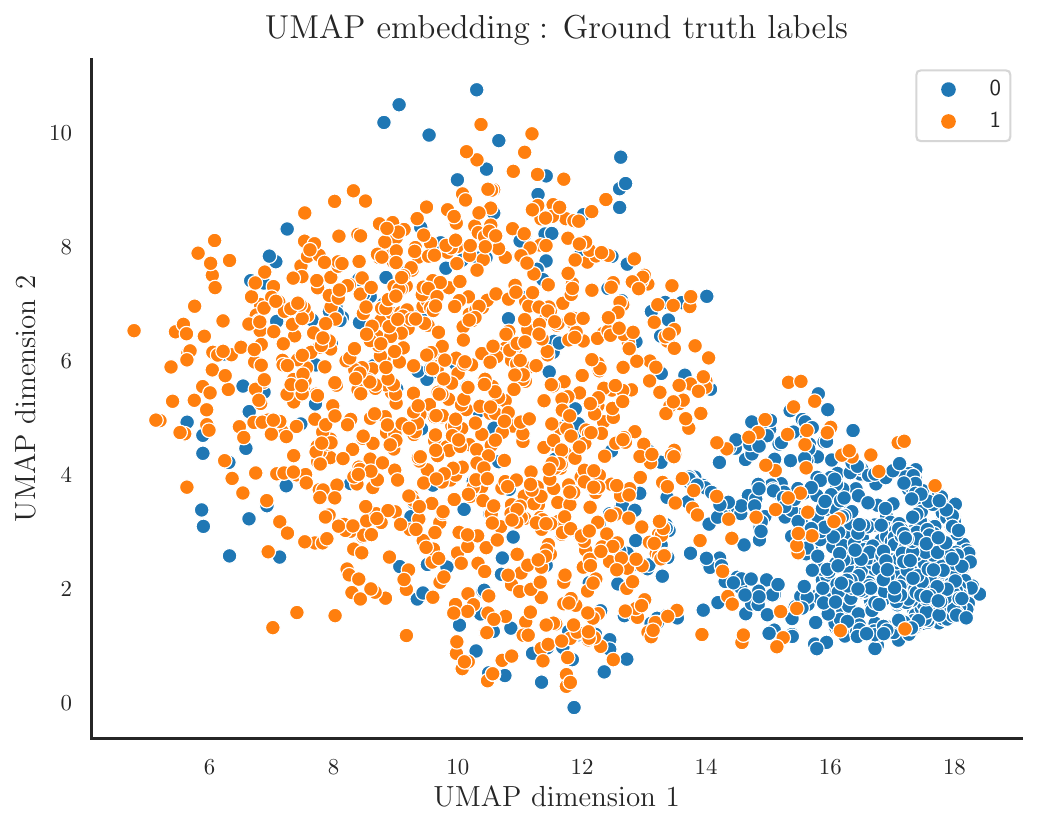} \\
\includegraphics[width=\linewidth]{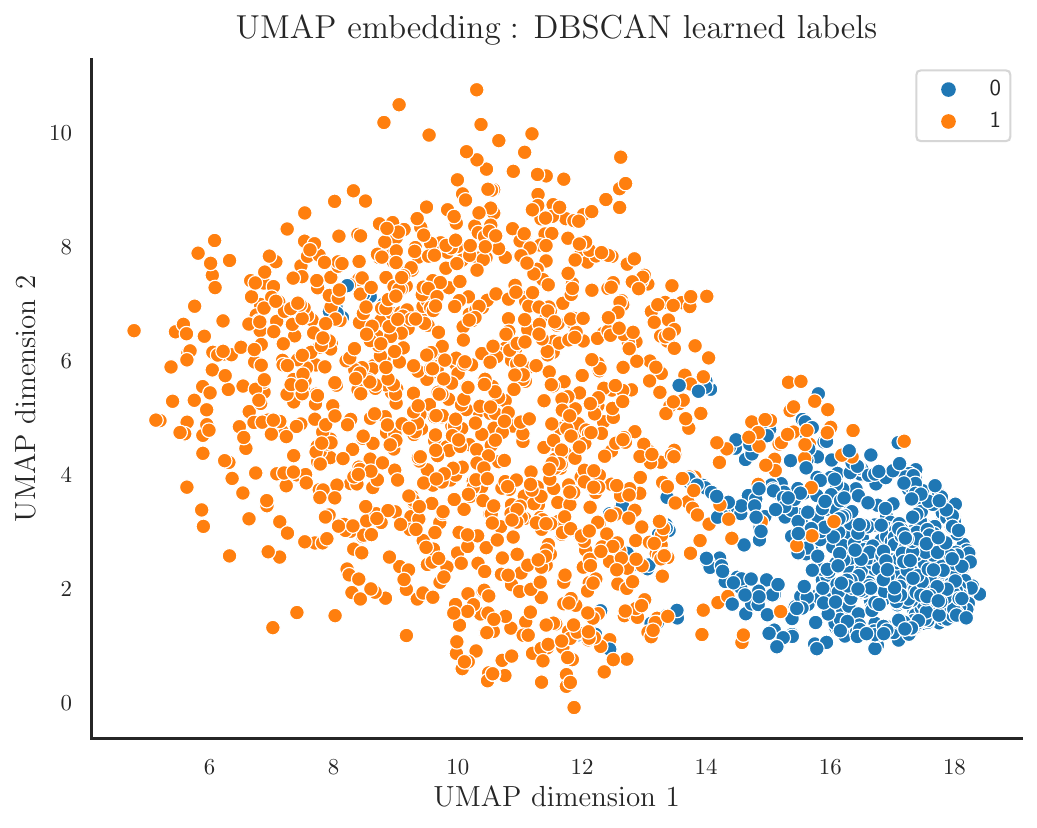} 
  \caption{UMAP embeddings of the data set colored according to ground truth labels (above) and their learned labels by DBSCAN (below)---QCD (0; blue) and top (1; orange).}
\label{fig:true-learned}
\end{figure}

We can visualize the efficacy of the clustering task through the embedding of the EMD metric space in a two-dimensional space obtained through use of the dimensionality-reduction algorithm UMAP \cite{sainburg2021parametric}. In our initialization of UMAP, we specify a denseMAP \cite{Narayan2020.05.12.077776} value 2.0 in order to preserve some of the local density profiles exhibited in the EMD space. Fig.~\ref{fig:true-learned} displays the UMAP embeddings of our data set labeled according to both their true labels, as well as their cluster assignments determined by DBSCAN.

We see that the vast majority of QCD jets are concentrated in a dense pocket in the lower right corner of the embedding, while only a couple handfuls or so are dispersed over the wide and sparsely-populated region dominated by the tops. Again, in light of our discussion in Sec.~\ref{sec:dbscan-business} and core/cluster point criterion defined by Eq.~(\ref{eq:dbscan-core}), we can understand why DBSCAN cannot extend the QCD cluster too far into the top/noise landscape.\footnote{Note the few blue points deep in the interior of the top region in the bottom plot of Fig.~\ref{fig:true-learned} are due the minor distance fluctuations/distortions inherent to the highly-nonlinear embedding function defined by UMAP. } Looking back to Fig.~\ref{fig:emds}, we see that the relative shift towards higher EMD values of the QCD-Top distribution relative to the QCD-QCD distribution is what allows for the density criteria to be satisfied by rather pure sample of QCD jets with minimal top contamination. This shift is then visualized in Fig.~\ref{fig:emds} by the separation of each class's effective center-of-mass in UMAP space. 

We remark that the physical origins of such a shift can be traced back to the presence of the additional scale ($m_{\mathrm{top}}$) characterizing the top jets which is absent in the QCD jet sample. That scale not only imprints itself upon the Top-Top EMD distribution, but more importantly does so almost equivalently on the QCD-Top distribution as well. Thus, one may reasonably inquire about whether such a feature is general to jets stemming from the decay of heavy-resonances relative to approximately scale-free QCD jets. We leave such interesting investigations to future work.

\section{Ricci flow on graphs}
\label{sec:ricci}

\subsection{Overview and graph-theoretic preliminaries}

In this section, we provide an alternative method for the unsupervised clustering of QCD and top jets that, like the previous section, utilizes only the geometric information induced by the EMD. We use the algorithm developed in \cite{ni2019community} for the unsupervised clustering of subgraphs (or ``communities'') within the graph defining the data set. The algorithm does this by computing Ricci curvatures between nodes in the graph and then iteratively updating the weights between nodes according to Ricci flow. Loosely speaking, Ricci flow works to compress and stretch regions of positive and negative curvature, respectively. The intuition is then that clusters within the graph, which are connected to one another through regions of negative curvature, can be stretched out and separated as distinct communities. We will outline key technical details behind the algorithm in what follows. 

First we must get our data set into a form that is amenable for this Ricci flow process. Doing so first requires a re-interpretation of our metric space $M = (D, \mathrm{EMD})$ as a weighted graph $G$ \cite{Diestel:2017}. This can be done in a very straightforward fashion by mapping $M$ to a fully-connected graph $G$ where each jet is represented by a node, with each node having edges connecting it to every other node, and then assigning edge weights given by the EMD distance between jets. The problem with such a procedure, though, is that it results in a fully-connected graph, but the Ricci curvatures for fully-connected graphs are uniformly positive---thus there will exist no regions of negative curvature for the Ricci flow algorithm to dilate and reveal any community structure among subgraphs. This is to say that we won't be able to resolve any clusters of jet type through Ricci flow following this procedure. 

However, there exists a simple way to induce regions of negative curvature at the graph level which relies on intuition gained in Secs.~ \ref{sec:emd-landscape} and \ref{sec:clustering}. Namely, we can make use of the $\kappa$-EMD defined in Eq.~(\ref{eq:kappa-def}) to reduce the edge set of a given vertex to define connections to a jet's $K$-nearest neighbors only, as opposed to the entirety of the data set $D$. Hence, we can consider each vertex to connect itself to a local patch of jets of a fixed cardinality, analogous the the core-point condition of DBSCAN, Eq.~(\ref{eq:core-point}), but substituting the ball-radius $\epsilon$ for $\kappa$, and $\mu$ for $K$, we will consider the analogue of a ball in a metric space for an edge set in a graph. Schematically,
\begin{align}
    |B_\epsilon (\jet) | \geq \mu \longleftrightarrow |B_\kappa (v_i)| = K\,.
\end{align}
The global connectivity of the graph is then induced via mutual overlap between neighboring patches about vertices. The remainder of this section will be devoted to building up the mathematical machinery required to make this correspondence precise.

Consider our collection of jets of point-cloud data type and size $N+1$, as in Eq.~(\ref{eq:data-set}). The EMD furnishes us a metric on elements of $D$, i.e. $\mathrm{EMD}_{ij}$ provides a distance between jets $\mathcal{J}_i$ and $\mathcal{J}_j$. This pairing of the set $D$ with the EMD metric is what defines the metric space 
\begin{align}
    M = (D, \mathrm{EMD})\,.
\end{align}
Let us recast $M$ as a graph. A weighted graph $G$ is a collection of vertices $V$, edges $E$, and edge weights $\omega$ \cite{Diestel:2017}, i.e.
\begin{align}
    G = (V, E, \omega) \,.
\end{align}
As previously mentioned, a fully-connected graph is a graph in which for every vertex $v_i \in V$ there exists an edge $v_iv_j \in E$ connecting $v_i$ to $v_j$ for all $v_j \in V$, and it is this structure that we want to modify. As such we will construct a graph whose vertices are connected $K$ other vertices, where $K$ is not necessarily equal to $N$. Taking $K=N$ will give the fully-connected case but in what follows, we will take $K$ to be general, ultimately linking it to the $K$ that defines the $\kappa$-EMD.

To begin, let us consider each jet in our metric space as a vertex in our graph, ultimately endowing the vertex set with the same labeling scheme as our data set
\begin{align}
    V = \left\{v_i \, \big | \, i \in I \right\}\,. 
\end{align}
All that differs is just that we are no longer keeping track of the underlying point-cloud nature of each jet. Such information is used only for the computation of EMDs, which we will take as given and will be carried by the edge weights. At this point, we see trivially that there is a one-to-one correspondence between the sets $D \longleftrightarrow V$.

We will consider the mapping from the metric space $M$ to the graph $G$ to be carried out schematically by the function $\Gamma$
\begin{align}
    M \xrightarrow{\,\,\,\Gamma\,\,\,} G\,.
\end{align}
This relation is schematic, as $\Gamma$ will really be a pair of functions, one for vertices and the other for edge weights.

The simplest instance of this mapping, which we will denote as $\Gamma_{\rm vertex}$, provides the correspondence between jets $\jet \in D$ and vertices $v \in V$:
\begin{align}
\Gamma_\text{vertex} \colon  D &\longrightarrow V \nonumber \\
\jet_i &\longmapsto  v_i\,,
\end{align}
which is to say that the image of $D$ under $\Gamma_\text{vertex}$ is $V$:
\begin{align}
    V = \Gamma_\text{vertex}(D)\,.
\end{align}

Now, given our set of vertices, we will define the edge set for each vertex. In doing so, we will make use of the EMD-set for each jet, defined by Eq.~(\ref{eq:emd-set-i}), and the $\kappa$-EMD of Eq.~(\ref{eq:kappa-def})
\begin{align}
    E_K (v_i) \equiv \left\{v_iv_j \, \big | \, \mathrm{EMD}_i \leq \kappa_i \,,\, j \in I\,,\, j \neq i \right\}\,,
\end{align}
where we adorn the set with a subscript $K$ to emphasize that the set inherits dependence on $K$ through the $\kappa$-EMD constraint. This definition makes it so that each $v_i \in V$ is connected to only $K$ neighbors. As such, the degree of each vertex, that is, the cardinality of the edge set for each vertex is $K$: $\mathrm{deg}(v_i) \equiv |E_K(v_i)| = K$. A set which comes as a natural byproduct of $E_K(v_i)$ is neighborhood set of each vertex $v_i$:
\begin{align}
    \nu_K (v_i) \equiv \left\{ v_j \, \big| \, v_iv_j \in E_K(v_i) \right\}\,.
\end{align}
We will make use of this set in the following section where we describe the mechanics of the Ricci flow algorithm.

The total edge set for the $K$-connected graph is then the union of the edge sets of each vertex
\begin{align}
    E_K = \bigcup _{i = 0}^N E_K(v_i)\,.
\end{align}

Once the $K$-restricted edge set $E_K$ is determined, the weight set is simple to define, as there is naturally a one-to-one correspondence between $\omega \longleftrightarrow E_K$. This correspondence is then induced by what we will label as $\Gamma_{\rm weight}$:
\begin{align}
\Gamma_\text{weight} \colon  E_K &\longrightarrow \omega \\
v_iv_j &\longmapsto  \mathrm{EMD}_{ij} \,.
\end{align}
This is to say that $\omega$ is the image of $E_K$ under $\Gamma_{\rm weight}$:
\begin{align}
    \omega = \Gamma_{\rm weight}(E_K)\,.
\end{align}

With the aforementioned sets in hand, we define the $K$-connected graph $G_K$ to be the collection
\begin{align}
    G_K = \lb V, E_K, \omega \rb\,,
\end{align}
where again, we highlight the fact that the edge set carries dependence on the choice of nearest-neighbor connections $K$, which ultimately controls all the interesting connectivity properties of the graph.

\subsection{Graph Ricci flow background}
In this section we will give an overview of the what goes into the computation performed by the Ricci flow algorithm of \cite{ni2019community}, following their exposition closely. In the previous section, we described the mapping that takes the metric space $M$ to the graph $G_K$. We will schematically write the algorithm of \cite{ni2019community} to be a mapping $\mathcal{R}: G_K \rightarrow G_K^\prime$ where $G_K^\prime$ is essentially $G_K$, but with its set of edge weights modified. The ``flow'' aspect of the Ricci flow will consist of iterating $\mathcal{R}$ numerous times, resulting in a family of edge-weight-modified graphs $G_K^{(t)}$, where $t$ can be thought of as a discretized ``time'' parameter. The workflow from our original metric space $M$ to the final graph with QCD and top subgraphs separated into clusters proceeds then as
\begin{align}
    M \xrightarrow{\,\,\, \Gamma \,\,\,} G_K^{(0)}
    \xrightarrow{\,\,\, \mathcal{R} \,\,\,} G_K^{(1)}
    \xrightarrow{\,\,\, \mathcal{R} \,\,\,} \cdots
    \xrightarrow{\,\,\, \mathcal{R} \,\,\,} G_K^{(n)}\,.
\end{align}

To define the Ricci curvature between vertices, we must first define a metric on the graph. We will refer to such an object as $d_G^{(t)}$, which is to say that it will be updated through each iteration of $\mathcal{R}$. This metric will be a function of the edge weights, thus inheriting its time-dependence through that of $\omega^{(t)}$. The metric for a general time $t$ is defined to be
\begin{align}
\label{eq:path-metric}
    d_G^{(t)}\left( v_i, v_j \right) =& \min \biggl\{ \sum_{k=0}^{n-1}\omega^{(t)}\lb v_{A_k}v_{A_{k+1}}  \rb \, \bigg| \,   \nonumber \\
    & v_{A_k}v_{A_{k+1}} \in E_K \,,\, v_{A_0} = v_i\,,\, v_{A_n} = v_j \biggr\}\,.
\end{align}
This may be intuitively understood as follows: due to the connectivity of the graph, while not all vertices are directly-connected to one another by a single edge, they are connected through a series of edges which can be thought of as those being traversed in hopping from one vertex to the next. Thus, we consider the set of all possible sequences of edges, sum their corresponding weights, and take the minimum length resulting from such traversals. This being the case, we can naturally relate $d_G^{(0)}$ directly to the EMD, as 
\begin{align}
\omega^{(0)}\lb v_{A_k}v_{A_{k+1}} \rb = \mathrm{EMD}_{A_kA_{k+1}} \,,    
\end{align}
albeit through the non-trivial relation of Eq.~(\ref{eq:path-metric}), which is a direct result of the non-trivial connectivity structure of $G_K^{(0)}$.

The process of computing the Ricci curvature between vertices in a graph involves the addition of one more structure, that is a probability measure defined on the graph. This probability measure will depend on the underlying graph metric $d_G^{(t)}$ and will thus inherit the time-dependence therefrom. The authors of \cite{ni2019community} define the graph measure to be
\begin{align}
\mathcal{P}^{(t)}_{i}(v_j) = \begin{dcases}
        \alpha & \text{if } v_j = v_i \\
        \frac{1 - \alpha}{Z_i}\exp \left(-d_G^{(t)}\left( v_i, v_j \right)^\beta \right) & \text{if } v_j \in \nu_K\left(v_i \right) \\
        0 & \text{otherwise }\,,
    \end{dcases}
\end{align}
where
\begin{align}
\label{eq:partition-fn}
    Z_i \equiv \sum_{v_j \in \nu_K (v_i)} \exp \left(-d_G\left( v_i, v_j \right)^\beta \right)\,,
\end{align}
is the effective partition function for the $i^{\rm th}$ vertex, providing normalization, and the parameter set $(\alpha, \beta)$ is taken to be $(1/2,2)$. The natural metric used for the space of measures is the 1-Wassertein distance $W_1$, whose formulation falls under the scope of optimal transport \cite{Villani:2009}, just like the EMD \cite{Komiske:2019fks, Komiske:2020qhg}. In doing so, we introduce a new metric space $(\mathcal{P}, W_1)$ atop the underlying graph metric space $(G_K, d_G)$. This pair allows us to then compute the Ricci curvature is between vertices $v_i$ and $v_j$ according to
\begin{align}
    R^{(t)}(v_iv_j)= 1 - \frac{W_1\left(\mathcal{P}^{(t)}_i, \mathcal{P}^{(t)}_j \right)}{d_G^{(t)}\left( v_i, v_j \right)}\,,
\end{align}
where the curvature is interpreted to be along the shortest path between $v_i$ and $v_j$. The curvatures then define the update edge weights---the flow then proceeds iteratively as
\begin{align}
    \omega ^{(t+1)} \left( v_iv_j \right) = \left(1 - R^{(t)}(v_iv_j) \right) \times d_G^{(t)}\left( v_i, v_j \right)\,.
\end{align}
For more details regarding the workings of the algorithm, see \cite{ni2019community}.

At this point, it is important to comment on the interpretation of the graph resulting from the process of Ricci flow. As stated previously, only the initial graph $G_K^{(0)}$ has direct relation to the physical EMD, that is has edge weights that can be directly related to angular resolution scale through which jets differ in their physical substructures. Of course, even at the level of $G_K^{(0)}$, while this physical interpretation is valid for each vertex in relation to its $K$-nearest neighbors, the modified connectivity structure blurs this meaning for distant jets that are connected via paths that require summing over many edges. For subsequent times, the edge weights $\omega^{(t)}$ completely lose their physical interpretation, as they deform according the will of the flow algorithm. As the the metric space defined by the EMD has a rich manifold structure \cite{Komiske:2020qhg}, it would be very interesting to understand any physical interpretation of its evolution through Ricci flow. This of course is far beyond the scope of the current work, which is simply to leverage geometric features in order to tag top jets from QCD. As such we leave such investigations to future work.

\begin{widetext}

\begin{figure}
\centering 
\includegraphics[width=0.35\linewidth]{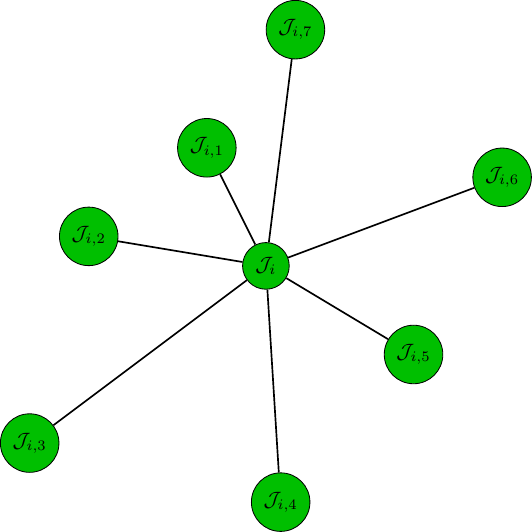}
\hspace{0.5cm}
\includegraphics[width=0.35\linewidth]{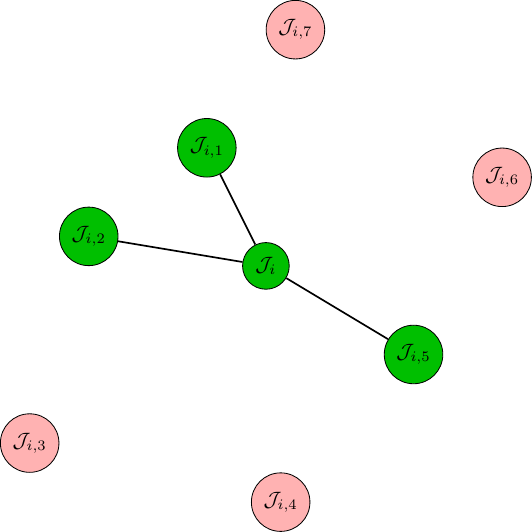} 
\caption{Schematic depiction of the modification to the connection structure for a hypothetical jet $\jet_i$ to its seven nearest-neighbors after reducing its connectivity to be that to its three. Note that only connections emanating from $\jet_i$ are depicted, the connectivity structures of its neighboring points are omitted for clarity.}
\label{fig:connectivity-example}
\end{figure}

\end{widetext}

\subsection{Top-tagging results from Ricci flow}

From the discussion of the previous section, we see that there is essentially only one parameter having to do with the structure of our data set that we must determine, and that is the nearest-neighbor number $K$ defining the reduced-connectivity structure we impose on our EMD metric space to transform it to that of a graph---see Fig.~\ref{fig:connectivity-example} for a visualization of this process. We find that there exists a good deal of freedom in the choice of this value, but values of $\mathcal{O}(10)$ perform the best, which, perhaps not surprisingly, can be understood from our discussion in previous sections. 

We note that in Sec.~\ref{sec:dbscan-parameters}, the determination of $\epsilon_{\mathrm{crit}}$ was due to it marking an extremum in the Shannon entropy of the cardinality distributions. Fig.~\ref{fig:eps-min-samples} depicts the cardinality distribution affiliated with the ball radius of $\epsilon_{\mathrm{crit}}$. We also know by Fig.~\ref{fig:emds} that this value of $\epsilon_{\mathrm{crit}}$ marks the peak of the QCD-QCD EMD distribution. Let us combine this information. Fig.~\ref{fig:emds} tells us that the majority of QCD jets are within $\epsilon_{\mathrm{crit}}$ of one another and Fig.~\ref{fig:eps-min-samples} corroborates this fact by telling us that the balls around most of these points contain sizeable portions of the QCD subset of the data, since $\mathcal{C} \sim 500$ for these balls. Thus, in choosing $K \sim \mathcal{O}(10)$,  while modifying the connectivity drastically (reducing the graph-ball cardinalities by an order of magnitude), the resultant effect on the graph-path metric between QCD jets is expected to be minimal due to the dense-packing of this corner of the metric space. One can conversely anticipate a more drastic change to in going from the EMD distances to the graph-path distances amongst tops, since their EMDs are Gaussian-distributed about $\theta_{\mathrm{cd}}$, so the reduction of their connection structure should amplify the number of ``$\theta_{\mathrm{cd}}$'s'' one must hop over to get from one corner of the top landscape to the other. This is all to say, that by reducing the connection structure drastically to $K \sim \mathcal{O}(10)$, we can exacerbate the graph-theoretic differences between QCD and top jets and use Ricci flow to amplify these differences. In the end, we choose $K=30$ based on these considerations, as $\mathcal{O}(1)$ multiples of this central value all yield similar results. We find that performing fifty iterations of Ricci flow is enough to achieve $\approx 91\%$ accuracy in our top-tagging task.

Interestingly, we find that there are two ways in which we can use the results from the Ricci flow algorithm in order to tag tops jets. The first of which is the most straightforward. This is the simply let Ricci flow run, updating the edge weights/graph-distances between points at each step. This works to effectively separate the portions of the metric space corresponding to QCD and top jets. To visualize this separation, we take the graph-path metric $d_G$, and embed its values between all elements of the data set into UMAP space, as done in Sec.~\ref{sec:clustering-results}.

\begin{widetext}

\begin{figure}[t!]
\begin{center}
\includegraphics[width=0.47\textwidth]{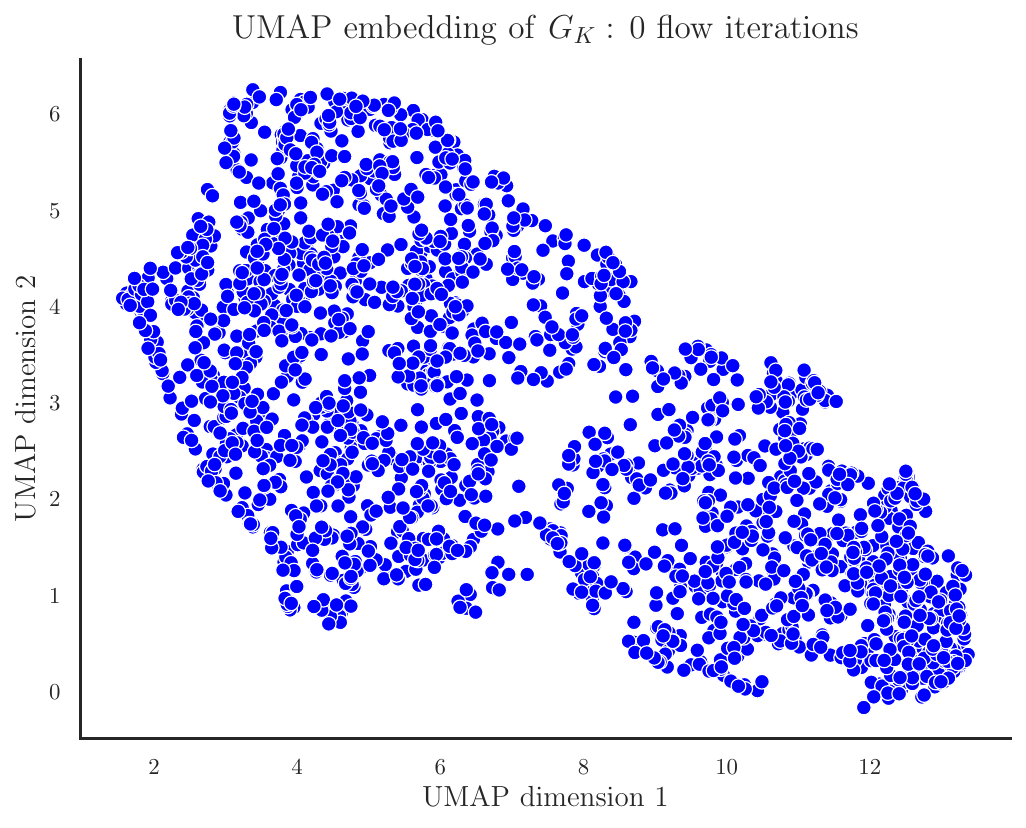}\hspace{0.1cm}
  \includegraphics[width=0.47\textwidth]{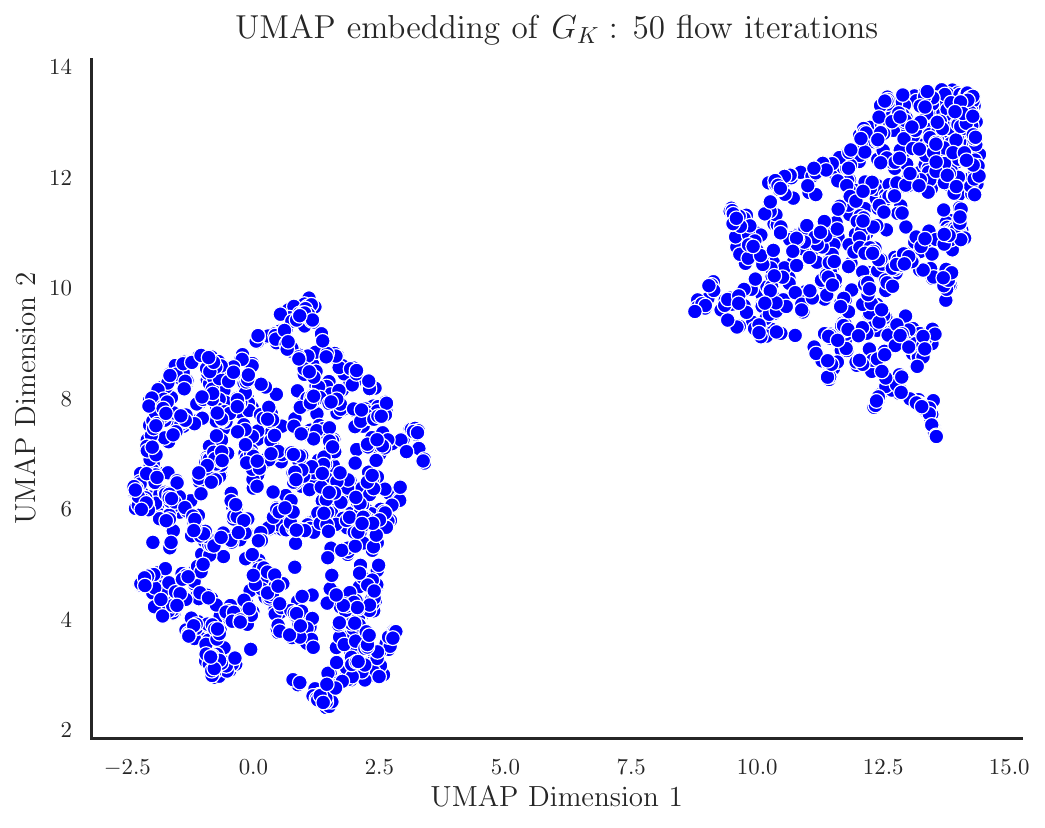} \\
\includegraphics[width=0.47\textwidth]{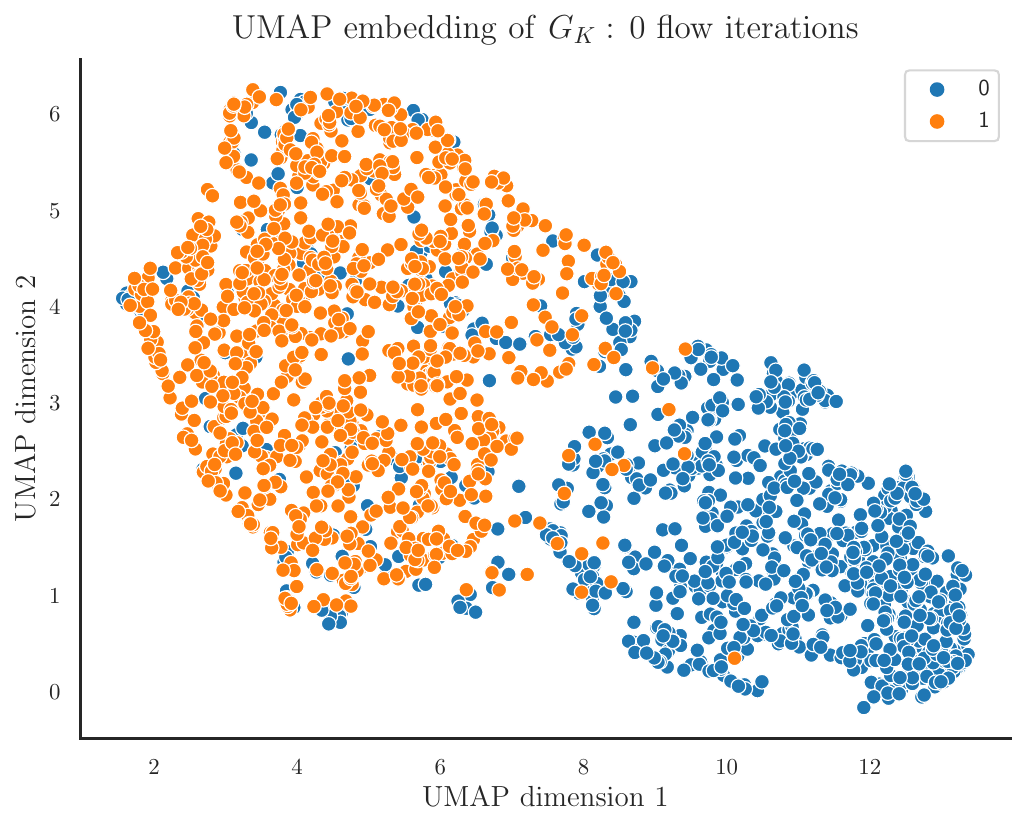}\hspace{0.1cm}
  \includegraphics[width=0.47\textwidth]{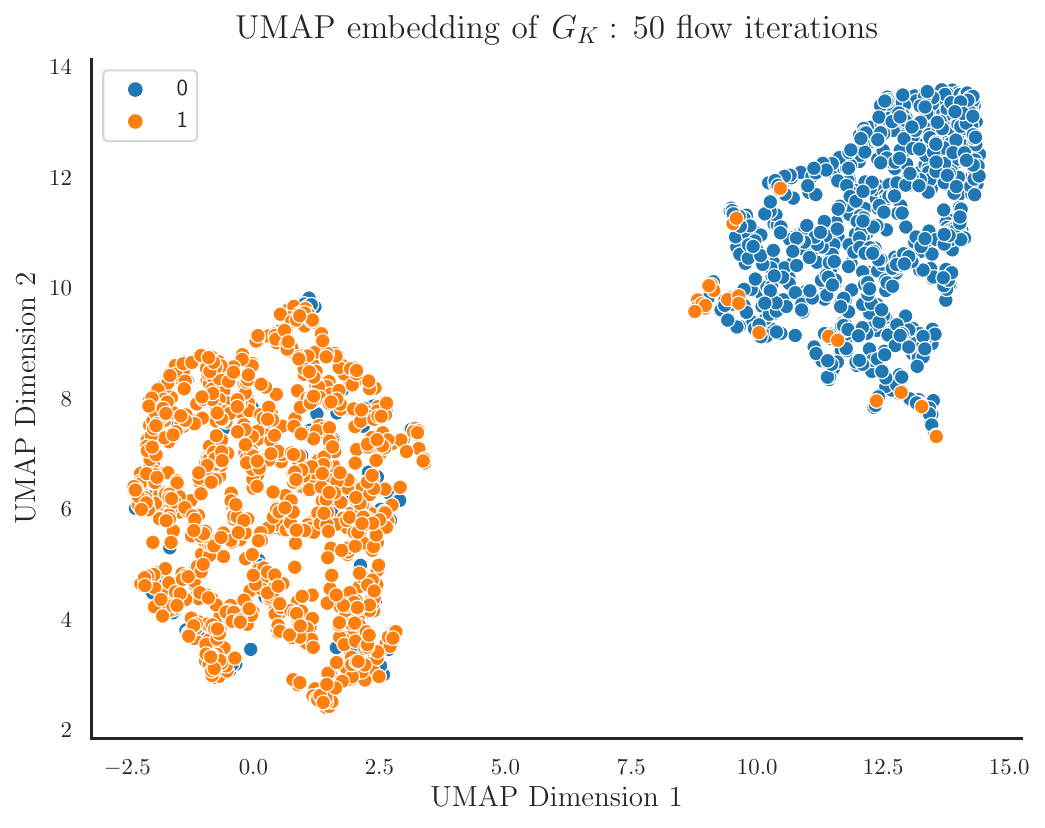}
\end{center}
  \caption{UMAP embeddings of the graph-path metric after zero (left) and fifty (right) iterations of Ricci flow. Results shown for unlabeled data above, with ground truth labels depicted below.}
\label{fig:umap-ricci}
 \end{figure} 

\end{widetext}

We depict such embeddings in Fig.~\ref{fig:umap-ricci}, where the top two panels show the embedding after zero (left) and fifty (right) iterations without labels, while the bottom two panels are labeled analogues. We see such a manifest separation of clusters in the UMAP embedding that one can perform a simple cut on the coordinates to ascribe labels. In doing so, we yield an accuracy of $91.04\%$. 

Interestingly, we can achieve similar accuracy through performing a cut on another output of the Ricci flow algorithm. Consider the following object, which can be thought of as analogous to the effective partition function in Eq.~(\ref{eq:partition-fn}):
\begin{align}
    \overline{R}^{(t)}(v_i) = \frac{1}{\mathrm{deg}(v_i)}\sum_{v_j \in \nu_K(v_i)} R^{(t)}(v_i v_j)\,.
\end{align}
This is effectively a Ricci curvature defined for an individual vertex, obtained through averaging over the curvatures between its $K$ nearest-neighbors. Thus the distribution of vertices induces a distribution over $\overline{R}$. We denote such distributions by $p_f\lb \overline{R} ; n \rb$, where $f$ is the flavor label and $n$ is the time step. We display such distributions in Fig.~\ref{fig:ricci-histos}, where the top two panels display the overall unlabeled distribution after zero (left) and fifty (right) iterations of Ricci flow, while the bottom two panels depict the underlying distributions colored according to their flavor. 

We see that the averaged-curvature distribution after fifty iterations of Ricci flow demonstrates a pronounced peak at the lower tail of its range, providing a clear value for a cut that separates the peak from the remainder of the distribution. In performing this cut and assessing the resultant accuracy, we achieve nearly the same level as with the UMAP cut: $91.13\%$.

What is particularly noteworthy about the distribution of averaged curvatures after fifty iterations of Ricci flow is the fact that the top jet distribution is strikingly localized while that of QCD jets is nearly-uniform. We can't help but posit that the latter is a reflection of the scale-free nature of QCD while the former has to do with the presence of the characteristic scale defined by $m_{\mathrm{top}}$. Of course, any direct numerical relation is obfuscated by the drastic modification that fifty flow iterations impact on the distances between jets, so any substantive conclusions are certainly beyond the scope of this work. We believe concerted effort to connect these concepts would be very interesting indeed.

\section{Discussion and conclusion}
\label{sec:conclusion}

In this work, we have found the metric space defined by the EMD to be rather rich with information. By analyzing the space defined by a samples of QCD- and top-initiated jets, we find jets of each class to fill their respective subspaces quite differently---in fact so differently that the labels of each jet can be determined purely through geometric information furnished by the EMD itself. The two methods pursued are (1) the density-based clustering through use of the DBSCAN algorithm and (2) the separation of subgraphs through use of the Ricci flow algorithm. Both of these methods are carried out completely unsupervised, achieve competitive accuracies, and each rely on only two initialization parameters. Furthermore, such parameters can be inferred from analyzing the data itself through physical reasoning. We compare some of the features to other leading top-taggers in Table~\ref{tab:compare}.

As can be seen, our methods distinguish themselves not only through their unsupervised nature, but also with the $\mathcal{O}(1)$ number of parameters required for initialization. As the sophistication of ML techniques applied to problems of jet substructure evolve, we believe that features such as simplicity and explainability are to become ever more important. As physicists, we don't only want tools that are effective in performing their tasks, but we also want tools whose effectiveness can be understood intuitively. Through our work, we see that such goals are certainly attainable and hope that the simple applications laid forth in this paper can serve as a starting point for more refined studies. One immediate extension would be to the unsupervised tagging of jets initiated by other heavy resonances, such as $W/Z/H$, as the presence of large mass scales in each case should presumably lead to a fair degree of separation from QCD jets in EMD space. It would be also interesting to analyze the circumstances needed for accurate clustering of jets in some low-dimensional space that the EMD manifolds are projected down to. In Ref.~\cite{Park:2022zov}, the authors developed a means of embedding manifolds whose distance is defined by the EMD into various two-dimensional subspaces in a way that preserves as many of the features of the true manifold as possible---e.g. ascribing directions corresponding to jet mass and pronginess. Carrying out a clustering analysis as a function of choice of two-dimensional embedding space would certainly be fascinating. We leave such investigations to future work.

\section*{Acknowledgements}
The authors are grateful to J.-H. Yang for his participation in the early stages of this work.  In addition, the authors would like to thank to Z.-B. Kang and A.J. Larkoski for helpful discussions. The work of JR is supported by the Mani L. Bhaumik Institute for Theoretical Physics.

\begin{widetext}

\begin{figure}[t!]
\begin{center}
\includegraphics[width=0.47\textwidth]{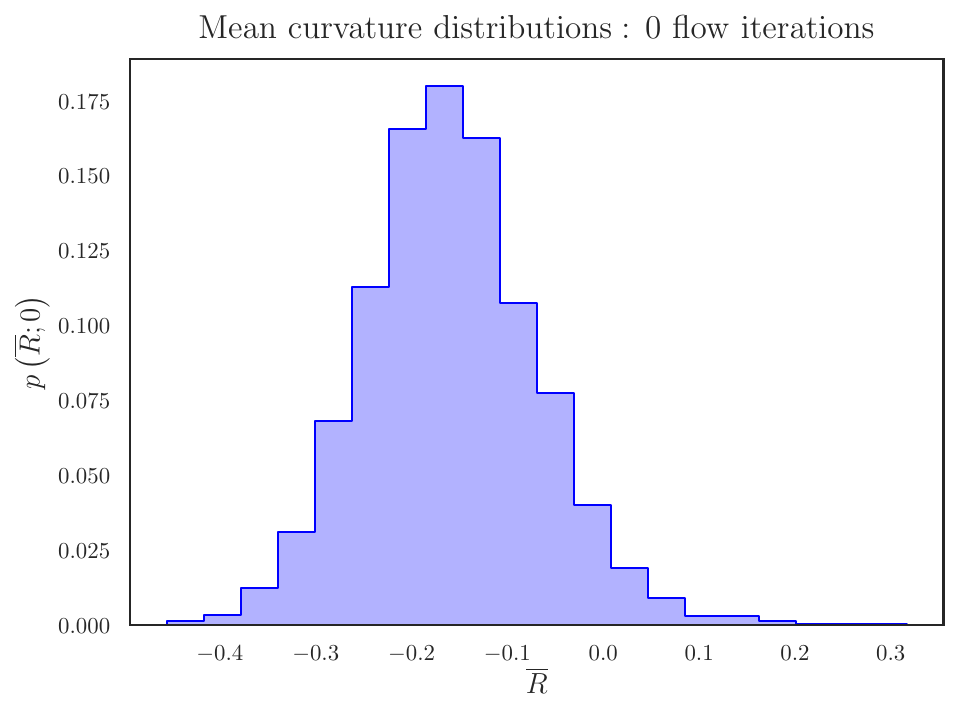}\hspace{0.1cm}
  \includegraphics[width=0.47\textwidth]{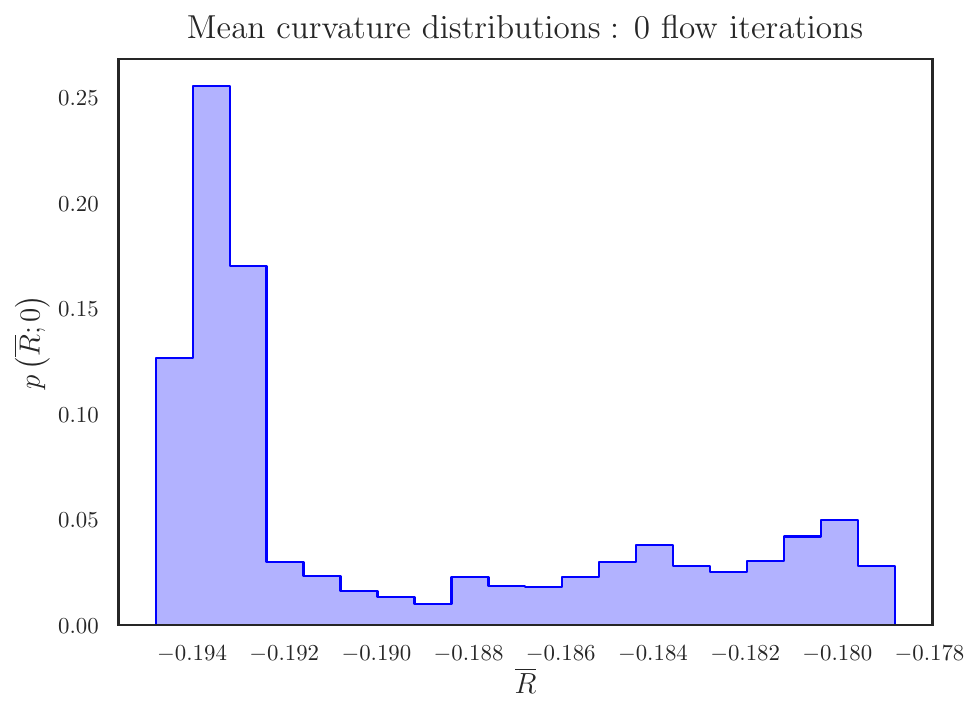} \\
\includegraphics[width=0.47\textwidth]{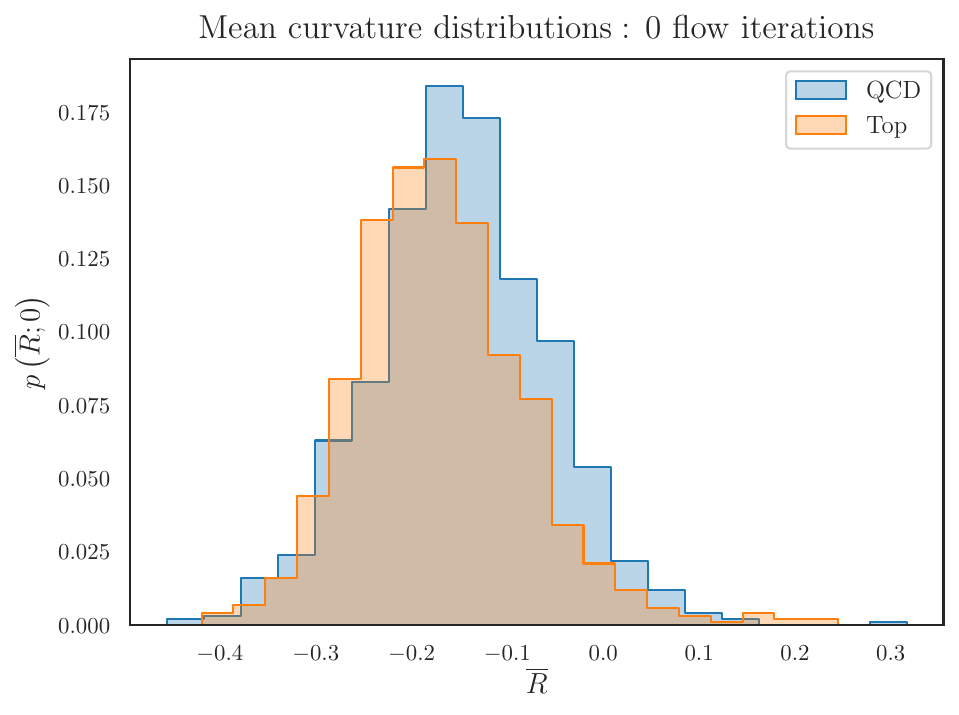}\hspace{0.1cm}
  \includegraphics[width=0.47\textwidth]{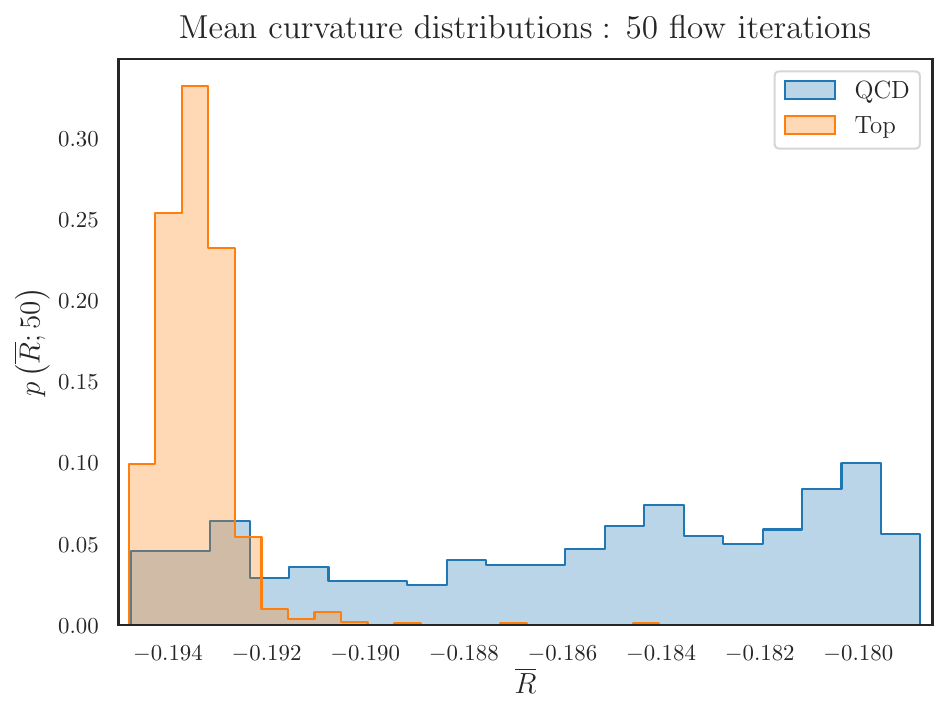}
\end{center}
  \caption{Averaged-curvature $\overline{R}$ distributions after zero (left) and fifty (right) iterations of Ricci flow. Results shown for unlabeled data above, with ground truth labels depicted below.}
\label{fig:ricci-histos}
 \end{figure} 
 


\begin{table}
\centering
\begin{tabular}{c | c | c | c }
\multicolumn{4}{c}{} \\
\hline
\hline
\textbf{Architecture} & \textbf{Accuracy} & \textbf{Parameters} & \textbf{Learning}\vspace{0.1cm} \\ 
\hline 
ResNeXt \cite{8100117}   & 0.9360 & 1.46e6 & Supervised \\
ParticleNET \cite{Qu_2020} & 0.9380 & 4.98e5 & Supervised \\
PFN \cite{Komiske:2018cqr} & 0.9320 & 8.20e4 & Supervised \\
LGN \cite{bogatskiy2020lorentz} & 0.9290 & 4.50e4 & Supervised \\
nPELICAN$_{\mathrm{hidden=1}}$ \cite{bogatskiy2023explainable} & 0.8951 & 11 & Supervised \\ 
DBSCAN$_{\mathrm{EMD}}$ & 0.9003 & 2 & Unsupervised \\
Ricci-Flow$_{\mathrm{Curvature}}$  & 0.9113 & 2 & Unsupervised \\
Ricci-Flow$_{\mathrm{UMAP}}$  & 0.9104 & 2 & Unsupervised \\
\hline
\hline
\end{tabular}
\caption{Comparison to a limited selection of top-taggers from the literature.}
\label{tab:compare}
\end{table}

\end{widetext}

\end{document}